\documentclass[aps,twocolumn,showpacs,superscriptaddress,citeautoscript,prl,floatfix]{revtex4-2}

\usepackage[utf8]{inputenc}

\usepackage{bm}
\usepackage{graphicx,float}
\usepackage{amsmath,amssymb}
\usepackage{amsfonts}
\usepackage{color}
\usepackage{color,soul}
\usepackage{physics}
\usepackage[normalem]{ulem}

\usepackage[dvipsnames]{xcolor}

\usepackage[colorlinks=true,%
            linkcolor=blue,%
            urlcolor=blue,%
            citecolor=blue,%
            filecolor=blue,% 
            bookmarksopen=true,%
            pdfauthor={DZ_JPRA_PAO},%
            pdftitle={MBS_BIC_MQDS},%
            pdfsubject={MBS_BIC_MQDS_Manuscript},%
            pdfpagemode=UseOutlines]{hyperref}
        
\definecolor{Blue}{rgb}{0.3,0.3,0.9}
\definecolor{Red}{rgb}{0.9,0.3,0.3}
\definecolor{Green}{rgb}{0.3,0.6,0.3}
\definecolor{Black}{rgb}{0.0,0.0,0.0}

\begin{document}

%\preprint{APS/123-QED}

\title{Observation of multi-orbital Fano resonances in photonic lattices}
%\title{Dynamical  Fano resonances in photonic lattices with multi-orbital impurities}% Force line breaks with \\

% Experimental realization of  the multi-orbital Fano.

\author{Diego Guzmán-Silva}
  \affiliation{Departamento de Física, Facultad de Ciencias Físicas y Matemáticas, Universidad de Chile, Santiago 8370448, Chile}
    \affiliation{Millennium Institute for Research in Optics - MIRO, Santiago, Chile}
\author{Maritza Ahumada}
\affiliation{Departamento de F\'isica, Universidad de Santiago de Chile (USACH), Avenida V\'ictor Jara 3493, 9170124, Santiago, Chile}

\author{Polette Parra-Palavecino}
 \affiliation{Departamento de Física, Facultad de Ciencias Físicas y Matemáticas, Universidad de Chile, Santiago 8370448, Chile}
    \affiliation{Millennium Institute for Research in Optics - MIRO, Santiago, Chile}

\author{Alexis R. Leg\'{o}n}
\affiliation{Departament of Physics, Universidad Técnica Federico Santa Mar\'ia, Valpara\'iso, Chile}
\author{Pedro A. Orellana}
\affiliation{Departament of Physics, Universidad Técnica Federico Santa Mar\'ia, Valpara\'iso, Chile}
\author{Rodrigo A. Vicencio}
    \affiliation{Departamento de Física, Facultad de Ciencias Físicas y Matemáticas, Universidad de Chile, Santiago 8370448, Chile}
    \affiliation{Millennium Institute for Research in Optics - MIRO, Santiago, Chile}

\date{\today}% It is always \today, today,

\begin{abstract}

%Fano resonances are such a fundamental physical phenomenon in which an open channel couples to a closed one, manifesting as a resonant cancellation of the transmission. In this article, we introduce a quantum simulator designed to observe multi-orbital Fano resonances based on a photonic waveguide analog. The system consists of a homogeneous lattice, including a plane wave generator, and an atom-like impurity that supports first-order (S) and second-order (P) modes. We derive a modified Fano model to describe this system and fit the experimental results based on this approach, demonstrating a remarkable agreement between experimental measurements and theoretical predictions. A full experimental characterization is performed, with clear double resonances, for S and P states, depending on the atom properties and the excitation wavelength. Our results show how the transport of a propagating beam on a lattice can be completely steered by managing the properties of an atom-like impurity, which acts as an efficient traffic light.

Fano resonances are a fundamental physical phenomenon that occurs when an open channel couples with a closed one, resulting in a resonant cancellation of transmission. In this article, we introduce a quantum simulator designed to observe multi-orbital Fano resonances using a photonic waveguide analog. The system consists of a homogeneous lattice, which includes a plane wave generator, and an atom-like impurity that supports first-order (S) and second-order (P) modes. 
We derive a two-impurity Fano-Anderson model to describe this system and fit the experimental results accordingly, demonstrating a remarkable agreement between our experimental measurements and theoretical predictions. A comprehensive experimental characterization reveals clear double resonances for the S and P states, which depend on the properties of the atom and the excitation wavelength. 
Our results illustrate how the transport of a propagating beam on a lattice can be effectively controlled by simply adjusting the properties of a single external atom-like impurity, which works as an efficient energy valve.
\end{abstract}

\maketitle

The Fano effect is one of the most distinctive interference phenomena observed in quantum mechanics. It was first introduced by Ugo Fano in 1961 within the field of atomic physics~\cite{Fano61}. Generally speaking, Fano resonances arise from the interference between a continuous band and a localized state, with a characteristic asymmetric line profile~\cite{FanoRMP}. A broad literature explores the Fano effect in various areas, including quantum transport, photonics, optics, and acoustics~\cite{Kobayashi,Fano-Opt,Fano-acous,Fano-QA}. Recent advances in nanotechnology, photonics, and metamaterials have led to numerous examples of resonant phenomena associated with Fano resonances. These findings have applications in fields such as switching and sensing~\cite{Fano-meta,Fano-Switch,Fano-sensing}. 
Additionally, the Fano effect has garnered significant attention, having been experimentally observed in various setups. These include quantum dots, optical waveguide arrays, cold atoms, and nonlinear photonic systems that utilize multiphoton scattering mechanisms, patterned dielectric slabs, dielectric gratings, cylinders, electromagnetic radiation, and arrays of nanoresonators~\cite{Naether2009,BECFano,Panov2024,Lishchuck2022,limonov2017fano}. Notably, the experimental observation of topologically protected Fano resonances was reported in acoustic systems, demonstrating increased robustness against disorder~\cite{TopoFano,ozawa2019}.

%Optical waveguide systems offer a clean and versatile platform for implementing a wide variety of configurations, thanks to micro- and nano-fabrication technologies that enable the creation of customized photonic structures. One of the key experimental techniques used to fabricate waveguide arrays is the femtosecond (fs) laser writing technique~\cite{szameit_discrete_2005,PRL2021Guzman}, which has demonstrated very fundamental physical phenomena~\cite{BICAlex,floquet2013,vicencio_observation_2015}. On the other hand, the experimental realization of controlled plane waves in physical systems is still a fundamental challenge; i.e., the generation of spatially extended wave profiles with a well-defined quasi-momentum distribution. For example, imaging techniques, based on spatial light modulators, are a way of modulating a wide optical beam in amplitude and phase. However, these techniques are optically limited to rather small patterns with an occupation of $\sim 7-9$ lattice sites and for very specific wavelengths only~\cite{CANTILLANO17}. However, recently, an innovative technique for generating wave packets with a narrow frequency distribution was demonstrated~\cite{Real2024}. There, the simple excitation of an edge site defect produces a Gaussian-like beam having a specific quasi-momentum distribution, which is controllable by the defect parameters.
Optical waveguide systems provide a clean and versatile platform for implementing a wide range of configurations, thanks to micro- and nano-fabrication technologies that enable the creation of customized photonic structures. One key experimental technique used to fabricate waveguide arrays is the femtosecond (fs) laser writing technique \cite{szameit_discrete_2005,PRL2021Guzman}, which has demonstrated fundamental physical phenomena as, for example, Bloch modes~\cite{Mandelik03}, Bloch oscillations~\cite{BOsc99}, Anderson localization~\cite{2DAnderson,1DAnderson}, PT-symmetric systems~\cite{LiebPTAlex}, edge topological states~\cite{floquet2013,stutzer_photonic_2018}, and Flat Band localization~\cite{vicencio_observation_2015,seba2015}, among others.

However, the experimental realization of controlled plane waves (PW) in physical systems remains a significant challenge, particularly the generation of spatially extended wave profiles with a well-defined quasi-momentum distribution. For instance, imaging techniques that utilize spatial light modulators can modulate a wide optical beam in terms of amplitude and phase. Nonetheless, these techniques are optically limited to relatively small patterns, typically occupying $\sim 7-9$ lattice sites, and for specific wavelengths only~\cite{CANTILLANO17}.
Recently, an innovative technique for generating wave packets with a narrow frequency distribution was demonstrated~\cite{Real2024}. In this method, the excitation of an edge site defect (wave packet generator) produces a Gaussian-like beam with a specific quasi-momentum distribution, which can be controlled by simply adjusting the defect parameters.

%\red{In this letter, we present the realization of a quantum simulator of multi-orbital Fano resonances in a photonic waveguide setup using a wave packet generation technique~\cite{Real2024}. Esto hay que trabajarlo al final!}
In this letter, we present a realization of a quantum simulator for multi-orbital Fano resonances, based on the direct correspondence between the paraxial wave equation in optics and the Schr\"odinger equation~\cite{JoannopoulosBook}. We develop a two-impurity Fano-Anderson model to describe the system and show a strong agreement between experimental measurements and theoretical predictions, confirming the validity of our model. A detailed experimental characterization reveals distinct double resonances for the $S$ and $P$ states, which depend on the properties of the atom-like defect and the excitation wavelength.
Our results show how a propagating beam in a lattice can be effectively controlled by only adjusting the properties of the atom-like impurity.

%Inset's upper panel: Most of the wave packet is transmitted through the lattice. Inset's lower panel: Most of the wave packet is reflected back to the input waveguide.

\begin{figure}
\includegraphics[width=\columnwidth]{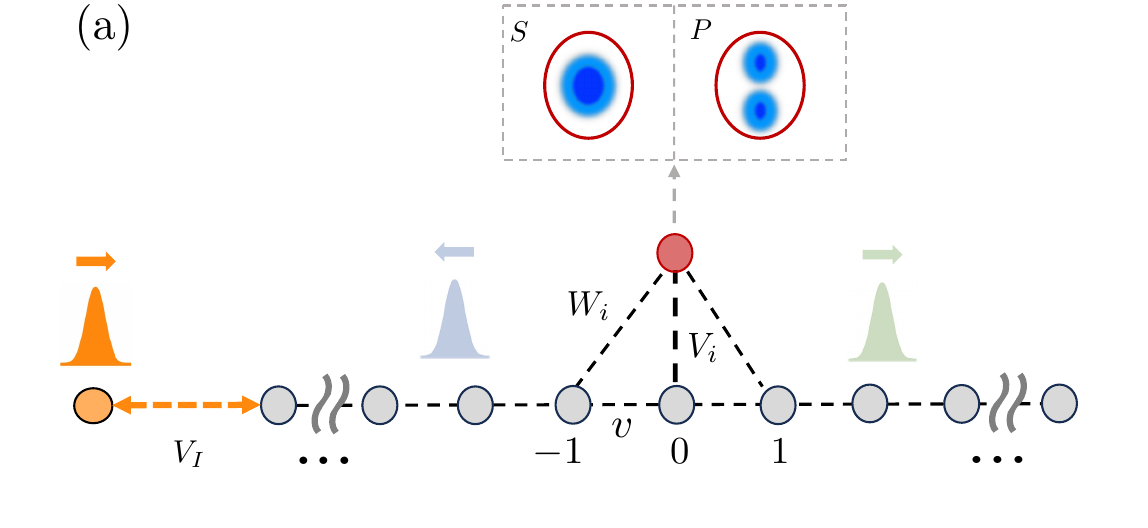}
\includegraphics[width=\columnwidth]{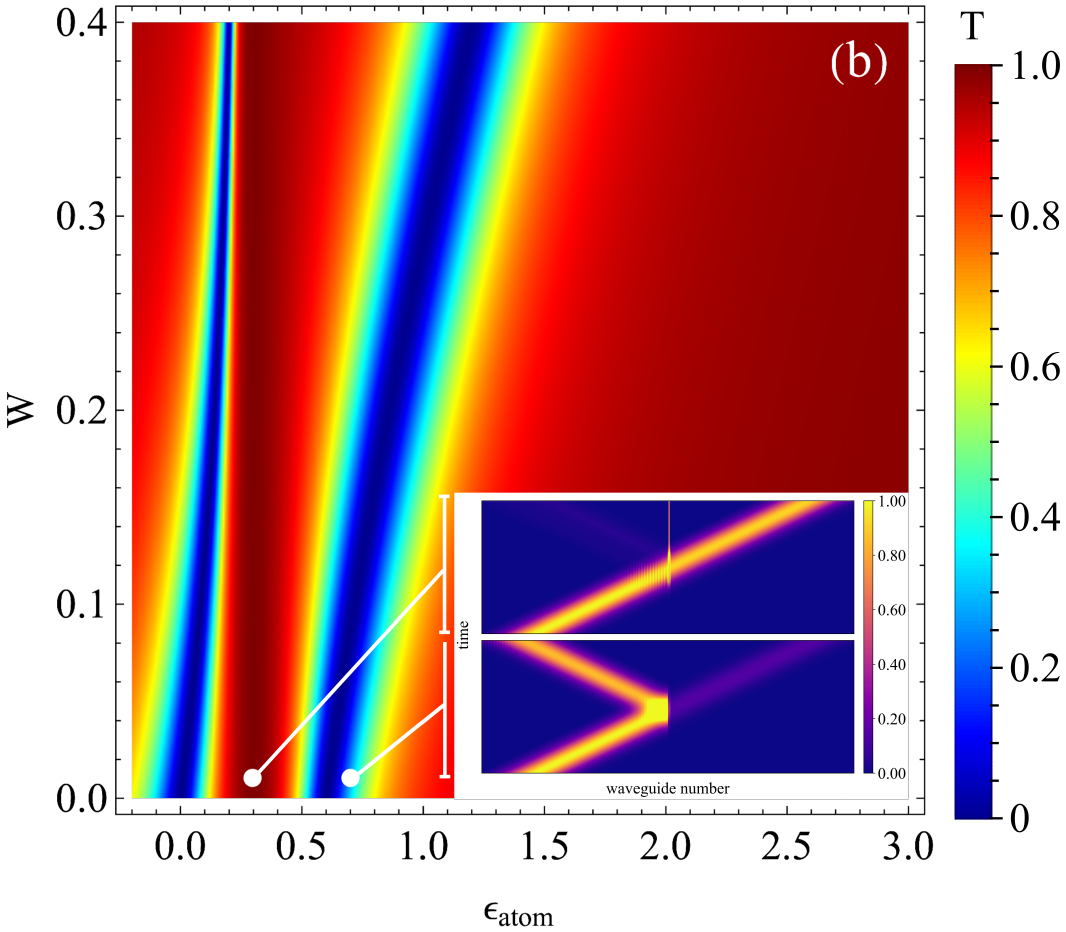}
\caption{Orange and red sites represent the wave packet generator and the atom-like impurity, respectively, the last supporting first-order ($S$) and second-order ($P$) modes. The orange dashed arrow indicates the coupling in between the lattice and the wave packet generator. Light green and light blue profiles represent the transmitted and reflected wave packets, respectively. (b) Transmission $T$ as a function of $\epsilon_{\rm{atom}}$ and NNN coupling $W_{S}=W_{P}=W$. Insets show dynamical probability profiles as a function of the waveguide number and time, for a Gaussian input excitation with $k_x=\pi/2$, and for $\epsilon_{\rm{atom}}=0.3$ (bottom) and $0.7$ (top) ($v=1$, $W=0.01$, $\omega_c=0$, $\epsilon_{S}=0$, $\epsilon_{P}=0.6$, and $V_{S}=V_{P}=0.5$).}
\label{Fig1}
\end{figure}

The system under investigation consists of a homogeneous one-dimensional (1D) lattice composed of weakly coupled optical waveguides and two impurity sites, as illustrated in Fig.~\ref{Fig1}(a). The impurity atom (red site) could support first ($S$) and second ($P$) order modes, depending on parameters. At the leftmost end of the system, a single defect (orange site) acts as a wave packet generator, introducing Gaussian-like propagating profiles having a well-defined quasi-momentum $k_{x}$~\cite{Real2024}. We model this problem by using a two-orbital Fano-Anderson approach, including also next-nearest-neighbor interactions at the ``atomic'' region [see diagonal dashed lines in Fig.~\ref{Fig1}(a)]. The Hamiltonian of this model (taking $\hbar=1$) reads as

\begin{eqnarray}\label{Hamilt}
H=\sum_{j}\left[\omega_{c}\hat{\Psi}_{j}^{\dagger}\hat{\Psi}_{j}-v(\hat{\Psi}^{\dagger}_j\hat{\Psi}_{j+1}+h.c.)\right]+\sum_{i=S,P}\Delta\varepsilon_{i}\hat{\Phi}_{i}^{\dagger}\hat{\Phi}_{i}\nonumber \\
-\sum_{i=S,P}\left[V_{i}(\hat{\Psi}_0^{\dagger} \hat{\Phi}_{i})+W_{i}(\hat{\Psi}_1^{\dagger} \hat{\Phi}_{i}+\hat{\Psi}_{-1}^{\dagger} \hat{\Phi}_{i})+\rm{h.c.}\right], \hspace{0.5cm}
\end{eqnarray}
% \begin{eqnarray}\label{Hamilt}
% H&=&\sum_{j}\omega_{c}\hat{\Psi}_{j}^{\dagger}\hat{\Psi}_{j}-\sum_{j}v(\hat{\Psi}^{\dagger}_j\hat{\Psi}_{j+1}+h.c.)\\ \nonumber
% &&+\sum_{i=S,P}\Delta\varepsilon_{i}\hat{\Phi}_{i}^{\dagger}\hat{\Phi}_{i}\,
% -\sum_{i=S,P}V_{i}(\hat{\Psi}_0^{\dagger} \hat{\Phi}_{i}+h.c.)\\ \nonumber
% &&-\sum_{i=S,P}W_{i}(\hat{\Psi}_1^{\dagger} \hat{\Phi}_{i}+\hat{\Psi}_{-1}^{\dagger} \hat{\Phi}_{i}+h.c.)\, ,
% \end{eqnarray}
%
where $\hat{\Psi}_{j}^{\dagger}$ ($\hat{\Psi}_{j}$) and $\hat{\Phi}_{i}^{\dagger}$ ($\hat{\Phi}_{i}$) are the creation (annihilation) operators of a single photon at the $j$-th lattice site and the impurity site, with frequencies $\omega_{c}$ and $\Delta\varepsilon_{i}$, respectively. 
We define the detuning as $\Delta\varepsilon_{i}=\epsilon_{i}-\epsilon_{\rm{atom}}$, where $\epsilon_{\rm{atom}}$ is a tunable parameter that can be swept to bring the system into or out of the resonance with the $i$-th impurity mode $\epsilon_i$.
%\red{Here, $\Delta\varepsilon_{i}=\epsilon_{\rm{atom}}-\epsilon_{i}$ defines the detuning between the atom energy and the energy of each impurity mode.} 
The parameter $v$ denotes the coupling strength between nearest-neighbor sites along the lattice. The coupling of the waveguide at site $j=0$ (below the atom) with the $S$ or $P$ atom modes is denoted by $V_{i}$. Additionally, $W_{i}$ defines the next-nearest-neighbor (NNN) coupling in between the atom state and sites $j=\pm 1$ in the lattice. The $S$ and $P$ modes do not interact at the atom impurity as they are orthogonal~\cite{PRL2021Guzman,APL25tutorial}. The dispersion relation of a 1D lattice is simply given by $E=\omega_{c}-2v\cos k_{x}$~\cite{rep1,Real2024}.
%%%%%%%%%
%%%%%%%%%%%%%%%%%%%%%%%%%%%%%%%%%%%%%%%%%%%%%%
%%%%%%%%%%%%%%%%%%%%%%%%%%%%%%%%%%%%%%%%%%%%%%

First of all, we theoretically study the scattering problem~\cite{FanoRMP,BECFano,Naether2009} by considering an incident plane wave traveling from the left (with an unitary amplitude), and a reflection $r$ and transmission $t$ resulting amplitudes.
We calculate the transmission probability defined as $T\equiv|t|^{2}$ by assuming that the incident wave-packet has a quasi-momentum $k_{x}=\pi/2$ (the faster PW in a 1D lattice~\cite{rep1}), obtaining 
\begin{equation}\label{Trans}
T = \frac{\mathcal{A}^{2} }{\mathcal{A}^{2} + \mathcal{B}^{2}}\ ,
%, \quad \text{where} 
\end{equation}
where $\mathcal{A}^{2} = 4 [(V_{S}W_{P}-V_{P}W_{S})^2 - 2v V_{S} W_{S}(\Delta_{P})- 2v V_{P} W_{P} (\Delta_{S}) - v^2 (\Delta_{P}) (\Delta_{S})]^2$ and $\mathcal{B}^{2} = v^2 \left[ V_{S}^2(\Delta_{P}) + V_{P}^2(\Delta_{S})\right]^2$. Here, $\Delta_{S}\equiv\epsilon_{S}-\epsilon_{\rm{atom}}-\omega_{c}$ and $\Delta_{P}\equiv\epsilon_{P}-\epsilon_{\rm{atom}}-\omega_{c}$. Equation~(\ref{Trans}) summarizes the scattering properties of the multi-orbital Fano effect, which can be characterized directly through experimental measurements of light intensity profiles (the details of the calculations are provided in Section I of the Supplemental Material~\cite{SM}). The theoretical expression for the transmission (\ref{Trans}) is plotted in Fig.~\ref{Fig1}(b), as a function of $W$ and $\epsilon_{\rm{atom}}$. We note that Eq.~(\ref{Trans}) exhibits zero-transmission points. The first one may occur for $\Delta_S=0$, but it is shifted by the term $(V_SW_P-V_PW_S)^2-2vV_SW_S\Delta_P$, corresponding to the Fano resonance associated with the first mode. A second null transmission occurs for $\Delta_P=0$, also shifted by $(V_SW_P-V_PW_S)^2-2vV_PW_P\Delta_S$. The last corresponds to a multi-orbital Fano resonance with the second mode of the atom. Both resonance shifts are clearly illustrated in Fig.~\ref{Fig1}(b).

The propagation of a given wave-packet through the system can be analyzed by means of the time-dependent Schr\"odinger equation $i\partial\psi/\partial z = H\psi$, where the coordinate $z$ plays the role of time in the experiments~\cite{rep1}. Numerical simulations of model (\ref{Hamilt}) are displayed in the insets of Fig.~\ref{Fig1}(b) (see section II at~\cite{SM}). In the upper inset, we observe a wave packet that does not resonate with the atom. In this case, a larger part of the energy is transmitted through the impurity region and the atom is effectively transparent to the wave packet. In contrast, when the atom impurity satisfies the resonance condition, a Fano resonance occurs and a nearly total wave packet reflection may be observed. Specifically, this happens for $\epsilon_{\rm{atom}}=\epsilon_S$ or $\epsilon_{\rm{atom}}=\epsilon_P$, with $\omega_c=0$ defining the zero energy point, as the bottom inset of Fig.~\ref{Fig1}(b) shows.

%In the following, we experimentally realize a quantum simulation of this system in a photonic waveguide setup.

%%%%%%%%%% FIGURA %%%%%%%%%%%%%%%%%%%
\begin{figure}[t!]
\includegraphics[width=1.0\columnwidth]{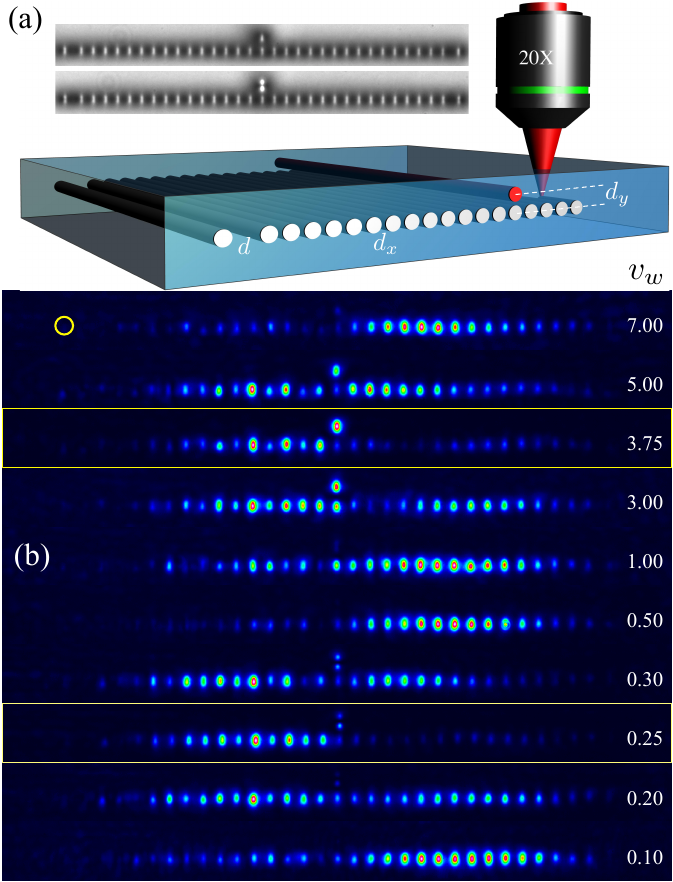}
\caption{(a) Sketch of the fs laser writing technique, with the atom waveguide colored in red. Insets: white-light microscope images of two fabricated arrays, the top (bottom) for an $S$ ($P$) atom waveguide. (b) Experimental intensity output profiles for an excitation wavelength of $670$ nm. The writing velocity $v_w$ (in mm/s) for the atom-like waveguide is indicated at the right. The yellow circle shows the input position, which generates a $\pi/2$ plane wave traveling to the right.}
\label{Fig2}
\end{figure}
%%%%%%%%%%%%%%%%%%%%%%%%%%%%%%%%%%%%%%
%\section{Experiments and results}

We experimentally implement the multi-orbital Fano-Anderson model by fabricating several photonic lattices using the femtosecond laser writing technique~\cite{szameit_discrete_2005}, as it is sketched in Fig.~\ref{Fig2}(a). A fs laser beam ($\lambda=1030$ nm and $\sim 215$ fs) is tightly focused inside a borosilicate glass wafer. The wafer, with a total length of $L=10$ cm, is translated along the propagation coordinate $z$ to create a single optical waveguide. This process is repeated for every waveguide in the lattice, generating photonic structures as the ones shown in Fig.~\ref{Fig2}(a)-insets. The system under study consists of three main components: the one-dimensional (1D) homogeneous waveguide array, the atom-like waveguide in the center above the lattice (see the red waveguide in the sketch), and the leftmost (or rightmost) PW generator edge site. The 1D lattice was fabricated with a fixed inter-site distance $d_x=17\ \mu$m, and a nominal coupling constant $v\sim 1.5$ cm$^{-1}$ at $730$ nm. The vertical distance for the ``atomic'' waveguide was set as $d_y=17.3\ \mu$m, such that $V_i\approx v$ (results for different values of $d_y$ are described at the Supplemental Material~\cite{SM}). The distance for the defect site was optimized to $d=22\ \mu$m, with $V_I\sim 0.7$ cm$^{-1}$ at $730$ nm. The defect waveguide was fabricated with the same lattice parameters, such to generate a PW with $k_x\approx \pi/2$~\cite{Real2024}, corresponding to the faster-propagating wave in 1D~\cite{rep1}.

The exploration of the parameter space where the Fano resonance is exhibited demands the adjustment of the atom-like waveguide properties; i.e., the tuning of the site energy $\epsilon_{\text{atom}}$, such to match the resonance condition described in Eq.~(\ref{Trans}). Experimentally speaking, this is achieved by modifying the refractive index contrast $\Delta n$ of the atom waveguide. Considering the fs writing technique, there are two different ways of performing this: by varying the writing power $P_w$~\cite{PRL2021Guzman} or by modifying the writing velocity $v_w$~\cite{Szameit_2010, Corrielli2013, Sebabrata2016}. Considering the need of having a smooth tuning process, we chose the second option, which also provides a wider range of $\Delta n$ in our experimental setup~\cite{SM}. The lattice and PW generator waveguides were fabricated with a nominal power of $P_w=30$ mW and a writing velocity of $v_w=5.0$ mm/s. A clear example of the atom waveguide detuning is shown in the insets of Fig.~\ref{Fig2}(a), by microscope images of two different lattices taken under white-light illumination. The $S$ (upper inset) and $P$ (lower inset) atom waveguides were fabricated for $v_w=5.0$ and $0.5$ mm/s, respectively. The multi-modal nature of the $P$ waveguide is evident in the lower inset.

We characterized our samples using a supercontinuum (SC) laser source, which enables us to sweep the excitation wavelength over a broad range, in this case $\lambda\in\{670,770 \}$ nm. We focus a horizontally polarized laser beam at the input facet of the glass chip, specifically at the PW defect site, and excite a propagating PW traveling to the right with $k_x\approx \pi/2$ (see Supplemental Material~\cite{SM} for more details). The variation of the excitation wavelength allows us to implement a wavelength-scan method~\cite{APL2023,Real2024}, which is equivalent to a dynamical propagation over the $z$ coordinate. As the mode width is proportional to the wavelength, the coupling constants increase linearly with $\lambda$~\cite{2DradPRL}, affecting directly the effective dynamical variable ``$v z$''.

Intensity images are captured by a beam profiler at the output facet of the glass chip, for the different fabricated lattices. An example for $\lambda=670$ nm is presented in Fig.~\ref{Fig2}(b), for some atom detunings controlled by different writing velocity at the atom waveguide. In this case, the plane wave is excited from the leftmost defect (indicated by a yellow circle), generating an extended Gaussian-like beam that travels to the right. We observe how for a lighter atom ($v_w=7.0$ mm/s), the plane wave has passed through the atomic region without much interaction, and with almost a full transmission. In this parameter region, and also for $v_w=0.5$ and $0.10$ mm/s, no interaction with the atom occurs. However, when the Fano resonance condition is satisfied, no PW transmission may occur beyond the position of the atom waveguide (right part of the structure), resulting in a reflected wave pattern. We observe this effect clearly at two different writing velocities in Fig.~\ref{Fig2}(b): first at $3.75$ mm/s, where a first mode is clearly observed at the atom position; and second, at $0.25$ mm/s where a second-order mode is visible in the atomic waveguide. This example provides a direct and concrete experimental proof of multi-orbital Fano resonances, where the transmission is abruptly diminished due to the interaction with an $S$ or a $P$ atomic state. We observe how the presence of a small and simple atomic disturbance can fully cancel the transmission of energy through the lattice. In addition, we observe a characteristic staggered profile for the reflected wavepacket, as it is expected for a $\pi/2$ incoming plane wave. When the Fano condition is not satisfied, we observe partially transmitted and partially reflected beams, as shown for $v_w=5.0$, $3.0$, $1.0$, $0.3$, and $0.2$ mm/s.

% %%%%%%%%%% FIGURA %%%%%%%%%%%%%%%%%%%
\begin{figure}[t!]
\includegraphics[width=1.0\columnwidth]{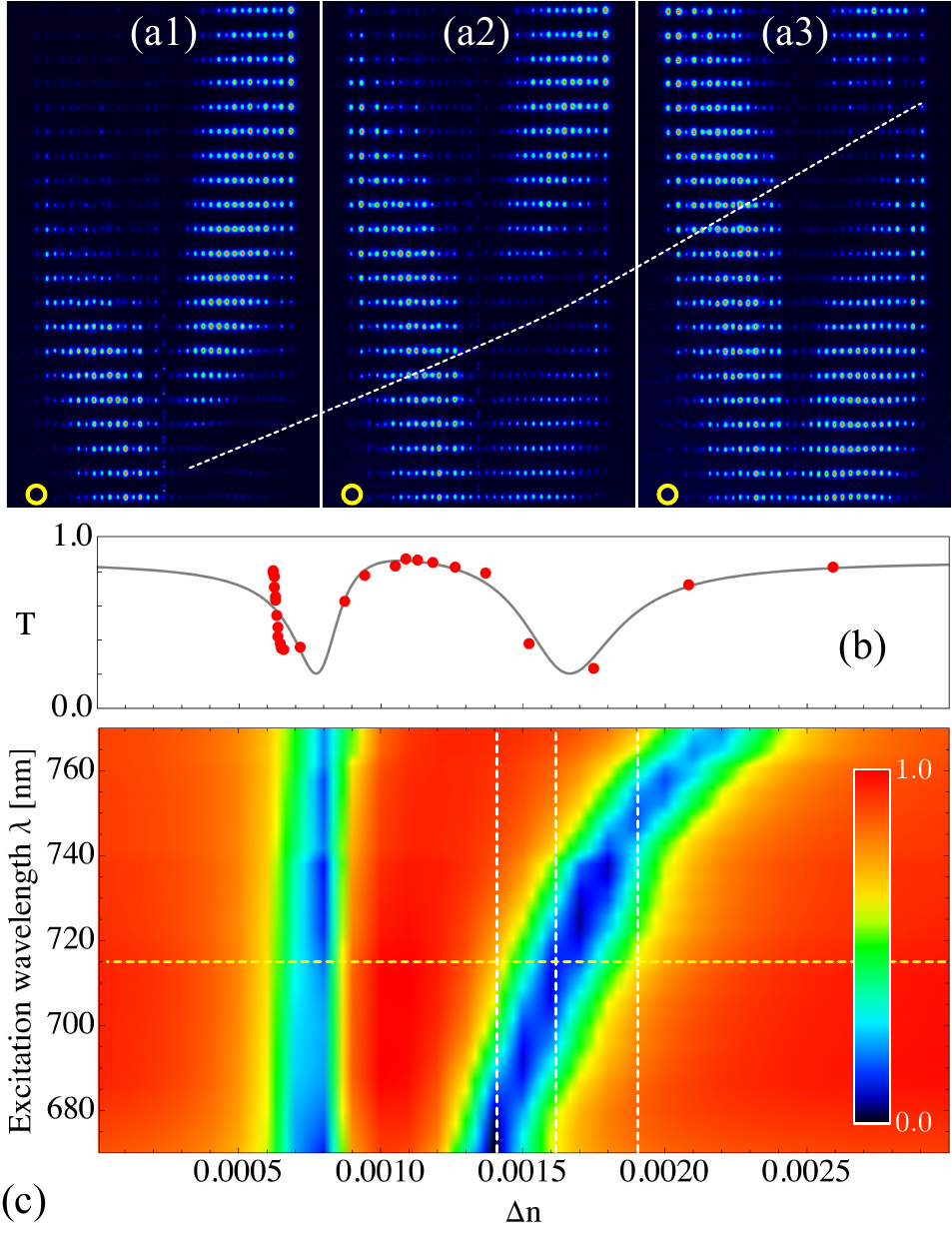}
\caption{(a) Output intensity profiles for wavelengths $670$ (bottom) to $770$ (top) nm, every 5 nm, for (a1) $v_w=0.30$, (a2) $0.25$, and (a3) $0.15$ mm/s. The yellow circle indicates the input position. (b) $T$ vs $\Delta n$ for $\lambda=715$ nm, with the experimental data shown by dots and the best-fit with a full line. (c) Density plot of the transmission $T$ as a function of the estimated $\Delta n$ and wavelength $\lambda$. Vertical and horizontal dashed lines correspond to the panels (a) and (b), respetively.}%\caption{Left panel: Density plot of the experimentally obtained transmission $T$ as a function of the calculated refraction index difference $\Delta n$ and the excitation wavelength $\lambda$, using the equation (\ref{ModTrans}). Right panel: $T$ vs $\Delta n$ for three different wavelengths, as indicated by the symbols (square, star, and circle correspond to $\lambda = 693$, $715$, and $738$ nm, respectively). Points represent the experimental data, and the continuous lines correspond to the best-fit curve.}
%the lines correspond to the best fits using (\ref{ModTrans}). \red{Disminuir labels. Poner `` wavelength $\lambda$ [nm]'' en vertical axis.}}
\label{Fig3}
\end{figure}
%%%%%%%%%%%%%%%%%%%%%%%%%%%%%%%%%%%%%%

The fabricated samples also allow us to analyze the shift in the resonances at different excitation wavelengths, which naturally depends on the atom detuning~\cite{SM}. Fig.~\ref{Fig3}(a) shows three examples where the reflected waves, originated from the Fano resonance with the $P$ state, are maximized at specific regions (see dashed line). For a weaker defect atom, we observe a reduction of the transmission for shorter wavelengths, while for stronger atoms the resonance occurs at larger values. This is because higher-order states emerge easily for shorter wavelengths~\cite{APL25tutorial} and, therefore, a $P$ resonance at larger values may occur for stronger (highlly detuned) atoms. Also, we notice that the wavelength resonant window becomes narrower for stronger atoms.

Finally, we perform a full wavelength-scan characterization of all the $25$ fabricated lattices, considering an incoming $k_x=\pi/2$ PW. For every lattice and every excitation wavelength, we obtain the probability transmission $T$, as a function of the atom refractive index contrast $\Delta n(\epsilon_{S,P})$, using a modified version of Eq.~(\ref{Trans}): $T=\alpha \mathcal{A}^2/(\mathcal{A}^2+\mathcal{B}^2)+\chi$,
%
% \begin{equation}\label{ModTrans}
% T=\alpha\frac{\mathcal{A}^2}{\mathcal{A}^2+\mathcal{B}^2}+\chi\ ,
% \end{equation}
%
where $\alpha$ and $\chi$ are fitting parameters. The conversion of $v_w$ for the atom waveguide into a refractive index contrast $\Delta n$ was performed by a direct characterization of the experimental guided modes and their comparison with realistic continuous simulations~\cite{SM}. Fig.~\ref{Fig3}(b) shows an example for $\lambda=715$ nm, where the dots correspond to the transmission obtained for every lattice. There, we can clearly observe the two Fano resonances. The full line is obtained after fitting the data with the modified model. The best agreement is found at the multi-orbital resonance for stronger atom waveguides (larger $\Delta n$), which is indeed the focus of this work. The collected results for all the lattices and for a fine wavelength-scan protocol are shown in Fig.~\ref{Fig3}(c). First of all, we notice that the first resonances (blue regions at smaller $\Delta n$) are shifted from the origin. In a standard Fano-Anderson model, the resonance must be exactly at $\Delta n=0$ for incoming $k_x=\pi/2$ plane waves. However, the experimental data shows a significant shift in the $S$ Fano resonance, which in our model is accounted by considering the coupling constants $V_i$ and $W_i$ of the atom waveguide with respect to the lattice~\cite{Miroshnichenko2005v2}. We also notice that this first resonance remains rather constant in $\Delta n$, for all the excitation wavelengths. On the other hand, the multi-orbital Fano resonance behaves differently, and the resonant value $\Delta n$ is observed to increase for larger wavelengths. This phenomenology is consistent with the results from Fig.~(\ref{Fig1}b), where the first resonance remains rather constant, but the second shifts its position for different values of the coupling and detuning.

In conclusion, we have proposed a lattice system where explicit Fano resonances can be observed directly by studying the wavefunction distribution in space. By introducing an additional defect waveguide, we were able to generate wave packets with a specific quasi-momentum around $k_x=\pi/2$, that interact with both the lattice and the atom-like waveguide. This interaction allows us to map the Fano resonance over a large range of parameters. We observe this phenomenon with two different modes, which can be considered as a multi-orbital Fano resonance process. The interaction at these higher modal levels has potential applications in scenarios that require on/off filtering at specific wavelengths, with an increasing selectivity for larger detunings. The experimental versatility of the photonic platform, together with its advanced wave packet generation technique, makes it suitable for simulating quantum systems in general. This proposal paves the way for developing quantum simulators capable of implementing a wide range of theoretical models that utilize, for example, plane-wave excitations as ignition mechanism. Finally, the addition of external circuitry on the top of the glass chip~\cite{extcontrol} could give extra degrees of freedom for achieving a completely controlled dynamics, which could transform into a very simple and concrete solution for photonic quantum devices.

% M.~A. acknowledges financial support from ANID Postdoctoral FONDECYT Grant No. 3240443. P.~A.~O. acknowledges support from DGIIE USM Grant PI-LIR-24-10 and FONDECYT Grants No. 122070 and 1230933. D.~G.-S. and R.~A.~V. acknowledge financial support from Millennium Science Initiative Program ICN17$\_$012. R.~A.~V. acknowledges financial support from FONDECYT Grant No. 1231313.

Authors acknowledge financial support from ANID Postdoctoral FONDECYT Grant No. 3240443, DGIIE USM Grant PI-LIR-24-10, FONDECYT Grants No. 122070, 1230933, and 1231313, and Millennium Science Initiative Program ICN17$\_$012.

\bibliography{FanoRefs,refs}% Produces the bibliography via BibTeX.

\end{document}